\documentclass[12pt,preprint]{aastex}
\shorttitle{Abundance of Isolated Clusters}
\shortauthors{Lee}
\begin{document}
\title{THE RELATIVE ABUNDANCE OF ISOLATED CLUSTERS AS A PROBE OF DARK 
ENERGY}
\author{Jounghun Lee}
\affil{Department of Physics and Astronomy, FPRD, Seoul National University, 
Seoul 151-747, Korea: jounghun@astro.snu.ac.kr} 
\begin{abstract}
Those galaxy clusters which do not belong to the superclusters are referred to as 
the isolated clusters. Their relative abundance at a given epoch may be a 
powerful constraint of the dark energy equation of state since it depends 
strongly on how fast the structures grow on the largest scale in the Universe. 
We note that the mass function of the isolated clusters can be separately 
evaluated through the modification of the recently developed Corasaniti-Achitouv 
(CA) theory according to which the stochastic collapse barrier is quantified by 
two coefficients: the drifting average coefficient ($\beta$) and the diffusion 
coefficient ($D_{B}$). Regarding $\beta$ in the CA formalism as an adjustable 
parameter and assuming that the formation of isolated clusters corresponds to the 
case of $D_{B}=0$, we determine the mass function of the isolated clusters by 
fitting the numerical results from the MICE simulations to the modified CA 
formula. It is found that the best-fit value of $\beta$ changes with redshift 
and that the CA mass function with $D_{B}=0$ agrees very well with the numerical 
results at various redshifts. Defining the relative abundance of the isolated 
clusters, $\xi_{I}$, as the ratio of the cumulative mass function of the isolated 
clusters to that of the non-isolated clusters at a given epoch, we finally 
show how sensitively $\xi_{I}$ changes with the dark energy equation of state.
It is also discussed how $\xi_{I}$ can help to break the degeneracy between 
the dark energy equation of state and the other key cosmological parameters. 
\end{abstract}

\section{INTRODUCTION}

The observed rich clusters of galaxies are believed to have formed through the 
gravitational collapse of the highest local peaks of the initial density field 
\citep{bbks86}. Although most of the rich clusters belong to the larger-scale 
superclusters \citep[e.g.,][]{wray-etal06,einasto-etal07}, there exist few 
isolated massive clusters which reside in relatively low-density environments  
without having any close neighbor clusters. The relative abundance of the isolated 
clusters should depend on how strong the gravitational clustering is on the 
largest scale and how frequently the clusters merge into the superclusters 
in the Universe. Therefore, it is expected that the relative abundance of the 
isolated massive clusters may be useful to distinguish between different 
candidate dark energy models.
   
To predict the relative abundance of isolated clusters in different dark energy models, 
however, a theoretical framework is first required within which the number density of 
isolated clusters can be separately counted. 
The standard excursion set formalism has been conventionally employed to 
analytically calculate the number density of galaxy clusters as a function of 
mass, i.e., the mass function of galaxy clusters. Basically, the standard 
excursion set formalism counts the initial regions whose linearly extrapolated 
density contrast ($\delta$) exceeds a fixed threshold value ($\delta_{c}$) on 
a certain mass scale $M$, regarding them as the sites which would eventually 
collapse to form the bound objects of mass $M$ \citep{PS74,bond-etal91}. 
The density threshold $\delta_{c}$ which is often called the {\it collapse 
barrier} has a constant value of $\delta_{sc}=1.686$ if the gravitational 
collapse process follows the spherical dynamics \citep{GG72}.

\citet{MR10a,MR10b} have for the first time introduced the concept of the 
stochastic collapse barrier to generalize the excursion set formalism. In the 
light of the numerical result of \citet{robertson-etal09}, \citet{MR10a,MR10b} 
regarded the collapse barrier, $\delta_{c}$, as a log-normal variable, and 
showed that in the high-mass limit the collapse barrier has a spherical 
average of $\langle\delta_{c}\rangle=\delta_{sc}=1.686$. 
Very recently, \citet{CA11a,CA11b} have made a further refinement of the mass 
function theory by incorporating the ellipsoidal collapse dynamics into the 
generalized excursion set formalism. They extended the generalized 
excursion set formalism to the case that the mean collapse barrier, 
$\langle\delta_{c}\rangle$, deviates from the spherical average, 
$\delta_{sc}=1.686$ in the low-mass limit. The higher collapse barrier than the 
spherical average, $\delta_{c}>1.686$, corresponds to the case that the 
formation of the low-mass halos  experience the disturbing tidal effect from 
the surroundings.

We note here that the Corasaniti-Achitouv formalism should be applicable not only 
to the case that the gravitational collapse occurs non-spherically in the 
low-mass section but also to the case that the massive halos form in 
isolated low-density environments. Given that the massive cluster-size halos 
are likely to form in the highly overdense regions but hardly in the isolated 
low-density regions, the collapse barrier for the formation of the isolated 
clusters should be higher than the spherical average. 
Since the Corasaniti-Achitouv formalism allows the average collapse 
barrier to deviate from the spherical value of $1.686$, their formalism 
may be capable of deriving the mass function of the isolated clusters. 

The organization of this paper is as follows. In \S 2 we briefly review the 
Corasaniti-Achitouv formalism of the halo mass function. In \S 3 we explain 
how the relative abundance of the isolated clusters can be evaluated by modifying 
the Corasaniti-Achitouv formalism and present the results of the numerical tests.  
In \S 4 we show how sensitively the relative abundance of the isolated clusters 
depends on the dark energy equation of state as well as the other key 
cosmological parameters.  \S 5 we summarize the results and conclude the work. 

\section{THE CORASANITI-ACHITOUV FORMALISM}

The  mass function of bound halos, $dN/d\ln M$, is defined as the 
number density of the bound objects whose masses belong to the differential mass 
interval $[\ln M,\ \ln M +d\ln M]$ \citep{PS74}. The classical mass function
theory based on the standard excursion set formalism relates $dN/d\ln M$ to the 
multiplicity function, $f(\sigma)$, which gives the number density of the 
random-walks crossing a specified collapse barrier, $\delta_c$, when the 
pseudo time variable has the value of $\sigma$ \citep{bond-etal91,jedamzik95}: 
\begin{equation}
\label{eqn:mft}
\frac{dN(M, z)}{d\ln M} = \frac{\bar{\rho}}{M}\frac{d\ln\sigma^{-1}}
{d\ln M}f[\sigma(M,z)].
\end{equation}
Here $\bar{\rho}$ is the mean mass density of the Universe and the pseudo 
time variable $\sigma(M,z)$ is equivalent to the rms density fluctuation of the 
linear density field smoothed on the mass scale $M$ at redshift $z$. It scales 
with the linear growth factor, $D(z)$, as $\sigma(M,z)\equiv D(z)\sigma(M)$ with 
$\sigma(M)\equiv \sigma(M,z=0)$. 
Throughout \S 2 and \S 3, we assume a flat $\Lambda$CDM cosmology 
(with $\Omega_{m}=0.25,\ \Omega_{\Lambda}=0.75,\ \Omega_{b}=0.045,\ h=0.7,\  
n_{s}=0.95$ and $\sigma_{8}=0.8$), for which the linear growth 
factor has the following analytical expression  \citep{lah-etal91}:
\begin{equation}
\label{eqn:D_lcdm}
D(z) \propto \frac{5}{2}\Omega_{m}
[\Omega_{m}(1+z)^{3}+\Omega_{\Lambda}]^{1/2}\int_{z}^{\infty}dz^{\prime}
\frac{1+z^{\prime}}{[\Omega_{m}(1+z^{\prime})^{3} + \Omega_{\Lambda}]^{3/2}}. 
\end{equation}
Here $D(z)$ is normalized to satisfy $D(z=0)=1$.

In the original Press-Schechter theory where the gravitational collapse is 
assumed to occur spherically, the collapse barrier has a constant value of 
$\delta_{sc}=1.686$. In the Press-Schechter variants developed afterward to 
improve the accuracy of the halo mass function by accounting for the ellipsoidal 
collapse process 
\citep[e.g.,][]{BM96,monaco97,audit-etal97,LS98,ST99,jenkins-etal01,
warren-etal06,tinker-etal08,robertson-etal09}, the collapse barrier is 
described not as a constant but as a function of the mass scale $M$ (or 
equivalently, a function of the rms density fluctuation, $\sigma$).
 
Recently, \citet[][hereafter MR10]{MR10a,MR10b} pointed out that not only the 
constant spherical collapse barrier but also the scale-dependent ellipsoidal 
collapse barrier is incapable of describing the true nature of the complicated 
halo formation process and suggested that the collapse barrier $\delta_{c}$ 
should be treated as a stochastic variable. MR10 derived an analytic 
expression for the halo mass function by generalizing the excursion set formalism 
for the case that the collapse barrier is stochastic and the random walk process 
is non-Markovian. It was shown by MR10 that in the high-mass limit 
($M\ge 10^{14}\,h^{-1}M_{\odot}$) the average of the stochastic collapse barrier 
equals the constant spherical value, $\langle\delta_{c}\rangle=\delta_{sc}=1.686$ 
and its variance is scale-dependent as 
$\langle\Delta^{2}D_{B}\rangle = \sigma^{2}(M)D_{B}$ 
where $D_{B}$ is called the diffusion coefficient. The stochasticity
of the collapse barrier is caused by the disturbance from the 
surroundings and the ambiguity in the halo-identification procedure 
\citep[see][for a detailed explanation]{MR10b}. 

In the light of the MR10 work, \citet[][hereafter CA11]{CA11a,CA11b} have 
extended the generalized excursion set formalism to the ellipsoidal collapse 
case in which the average of $\delta_{c}$ deviates from the spherical value, 
$1.686$, in the low-mass section. 
The resulting CA11 multiplicity function has two characteristic coefficients: 
the diffusion coefficient $D_{B}$ defined as in the 
MR10 formalism and the drifting average coefficient, $\beta$, defined as  
$\langle\delta_{c}\rangle \equiv \delta_{sc} + \beta$. The non-zero value 
of this drifting average coefficient, $\beta$, quantifies the degree of the 
deviation of the average of the collapse barrier from the spherical average 
$1.686$, while the non-zero value of the diffusion coefficient $D_{B}$ 
quantifies the degree of the stochasticity of the collapse barrier 
$\delta_{c}$.

The CA11 multiplicity function $f(\sigma)$ was written at second order as 
\begin{equation}
\label{eqn:ca10_mul}
f(\sigma ; D_{B},\beta) \approx f^{(0)}(\sigma ; D_{B}) + 
f^{(1)}_{\beta=0}(\sigma ; D_{B}) + 
f^{(1)}_{\beta}(\sigma ; D_{B}) + 
f^{(1)}_{\beta^2}(\sigma ; D_{B}),
\end{equation}
where 
\begin{eqnarray}
\label{eqn:f0}
f^{(0)}(\sigma;D_{B})&=&\frac{\delta_{sc}}{\sigma\sqrt{1+D_B}}
\sqrt{\frac{2}{\pi}}\,e^{-\frac{(\delta_{sc}+\beta\sigma^2)^2}
{2\sigma^2(1+D_B)}},\\
\label{eqn:f1b0}
f^{(1)}_{\beta=0}(\sigma;D_{B})&=&-\tilde{\kappa}\frac{\delta_{sc}}{\sigma}
\sqrt{\frac{2a}{\pi}}\left[e^{-\frac{a\delta_{sc}^2}{2\sigma^2}}
-\frac{1}{2}\Gamma\left(0,x\right)\right],\\
\label{eqn:f1b1}
f^{(1)}_{\beta}(\sigma;D_{B})&=&
-\beta\,a\,\delta_{sc}\left[f^{(1)}_{\beta=0}(\sigma)+\tilde{\kappa}\,
\textrm{Erfc}\left(x\right)\right],\\
\label{eqn:f1b2}
f^{(1)}_{\beta^2}(\sigma;D_{B})&=&\beta^{2}a^{2}\delta^{2}_{sc}\tilde{\kappa}
\biggl\{\textrm{Erfc}\left(x\right)+
\frac{\sigma}{a\delta_{sc}}\sqrt{\frac{a}{2\pi}}\biggl[e^{-\frac{a\delta_{sc}^2}
{2\sigma^2}}\left(\frac{1}{2}-2x\right)+\frac{3}{4}\frac{a\delta_{sc}^2}
{\sigma^2}\Gamma\left(0,x\right)\biggr]\biggr\},
\end{eqnarray}
with $x\equiv (\delta_{sc}/\sigma)\sqrt{a/2}$, $a=1/(1+D_{B})$, 
$\tilde{\kappa}=\kappa\,a$, $k=0.475$, and incomplete Gamma function 
$\Gamma(0,x)$.  Here, $f^{(0)}(\sigma;D_{B})$ coincides with the MR10 
multiplicity function which has only one parameter $D_{B}$. 
Computing the halo mass functions through Equations 
(\ref{eqn:mft})-(\ref{eqn:f1b2}) and comparing them with the fitting formula given 
by \citet{tinker-etal08}, CA11 have determined the best-fit values of the two 
coefficients as $\beta=0.057$ and $D_{B}=0.294$. 
CA11 have also shown how the shape of the halo  mass function changes with the 
values of $D_{B}$ and $\beta$ \citep[see Figure 3 in][]{CA11b}. 
According to their results, the variation of the diffusion coefficient 
$D_{B}$ alters the high-end slope of $f(\sigma)$ while the variation of the 
drifting average coefficient $\beta$ affects the over-all amplitude of 
$f(\sigma)$.

Now, we would like to confirm the validity of the CA11 mass function by testing 
it against the high-resolution MICE simulations \citep{mice10} which traced the 
evolution of $2048^{3}$ dark matter particles (each having mass of  
$23.42\times 10^{10}\,h^{-1}M_{\odot}$) on a periodic box with linear size 
of $3\,h^{-1}$Gpc for a $\Lambda$CDM cosmology. 
We utilize the publicly available catalog of the cluster halos that were 
identified via the standard Friends-of-Friends (FoF) algorithm. The catalog 
provides information on each halo's  (comoving) position, peculiar velocity and 
FoF mass (calculated as the sum of the masses of all the dark matter particles 
belonging to each halo) at three different redshifts: $z=0,\ 0.5,\ 1$. 
For a detailed description of the MICE simulations and the cluster catalogs, 
we refer the readers to \citet{mice10}. 

Binning the logarithmic masses of the cluster halos from the MICE catalog and 
counting the number densities of the cluster halos belonging to each logarithmic 
mass-bin, we obtain the numerical mass function of all cluster halos, 
$dN_{T}/d\ln M$, at each redshift. 
Fitting the CA11 formula (Eqs [\ref{eqn:mft}]-[\ref{eqn:f1b2}]) to the 
numerical results from the MICE simulations at each redshift, we have determined 
the best-fit values of $\beta$ and $D_{B}$ with the help of the 
$\chi^{2}$ statistics, the results of which are listed in Table \ref{tab:best1}. 

Note that the diffusion coefficient, $D_{B}$, has the same best-fit value of 
$0.38$ at all three redshifts while the best-fit value of the drifting average 
coefficient, $\beta$, increases as the redshift decreases. 
The result indicates that at earlier epochs the gravitational collapse process 
is closer to the spherical dynamics than at present epoch. 
Since a bound halo is harder to form at earlier epochs, those high-$z$ halos 
must correspond to higher peaks in the initial density field. As shown 
analytically by \citet{bernardeau94}, the higher a local density peak is, the 
more spherically its gravitational collapse proceeds.

It is also worth mentioning that our result on the best-fit value of 
$D_{B}=0.38$ is different from the original value $D_{B}=0.27$ used in the 
CA11 formalism.  This discrepancy must be due to the fact that we use the FoF 
masses available in the MICE catalog while the spherical 
over-density (SO) masses were considered in the original CA11 work.

\begin{figure}
\begin{center}
\plotone{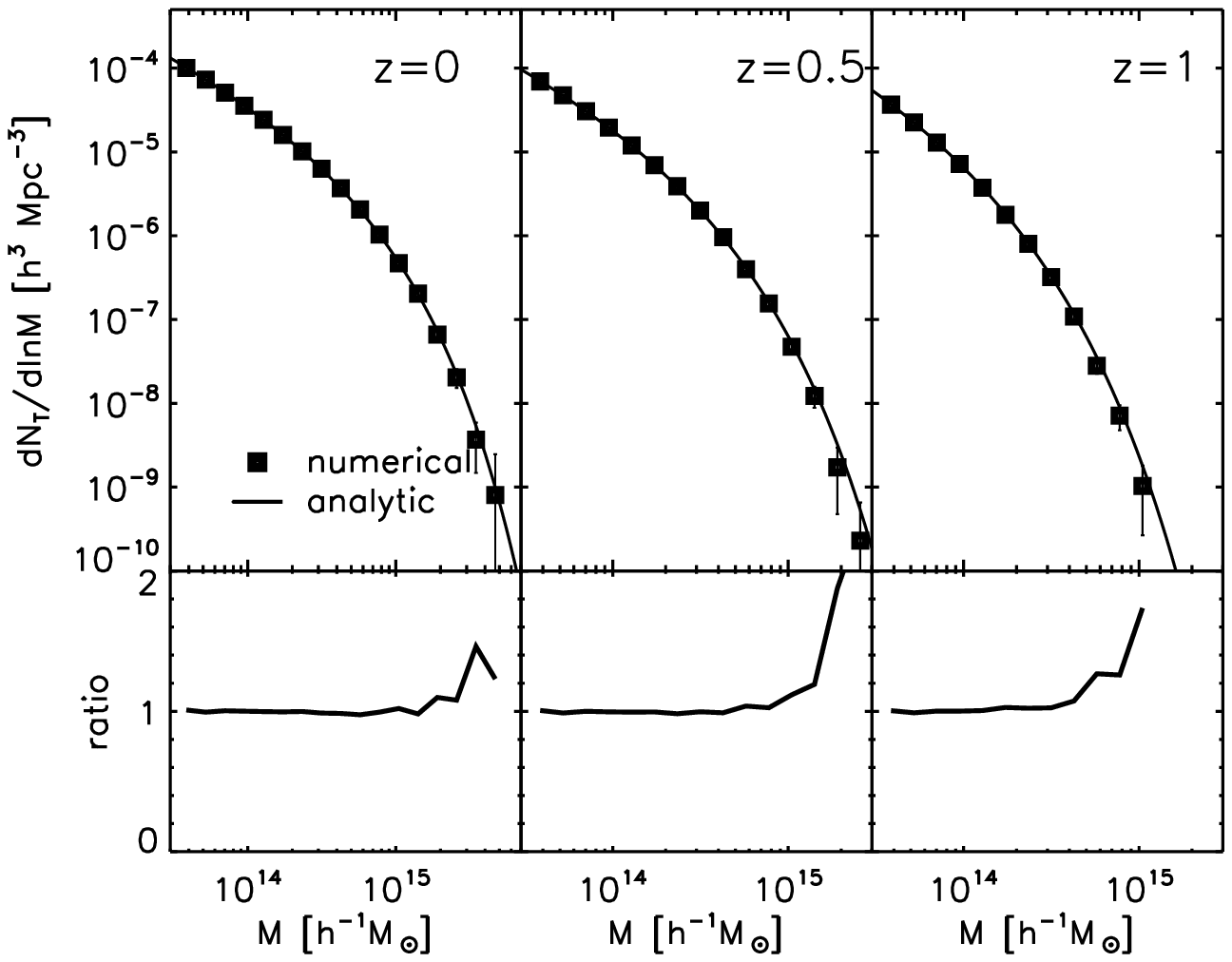}
\caption{Mass function of cluster halos (top panel) and the ratio of 
the analytic prediction to the numerical result (bottom panel) at $z=0,\ 0.5$ 
and $1$ (in the left, middle and right panels, respectively). In each of the 
top panels the square dots represent the numerical result from the MICE 
simulations while the solid line is the analytic prediction based on the CA 
formalism. The errors are the one sigma scatter among the eight Jackknife 
resamples.}
\label{fig:dndm}
\end{center}
\end{figure}

Figure \ref{fig:dndm} plots the CA11 mass functions with the best-fit values 
of $\beta$ and $D_{B}$ (solid line) and compare them with the numerical results 
from the MICE simulations (square dots) at $z=0,\ 0.5$ and $1$ in the top-left, 
top-middle and top-right panels, respectively. 
The Jackknife method is employed to calculate the errors: Dividing the halos 
into eight subsamples (each having the same number of the halos) and determining  
$dN_{T}/d\ln M$ separately from each subsample, the errors are calculated 
as one $\sigma$ scatter of $dN_{T}/d\ln M$ among the eight Jackknife 
resamples. Figure \ref{fig:dndm} also plots the ratio of the analytic mass 
function to the numerical result as a function of mass at $z=0,\ 0.5$ and $1$ 
in the bottom-left, bottom-middle and bottom-right panel, respectively.
As can be seen, the analytical and numerical mass functions agree with each other 
quite at all three redshifts, except in the high-mass section 
($M> 10^{15}\,h^{-1}M_{\odot}$) where the Jackknife errors are very large. 

\section{MASS FUNCTION OF THE ISOLATED CLUSTERS}

Applying the FoF algorithm with the linkage length parameter of $b$ to the MICE 
cluster catalogs at each redshift, we first find the clusters of clusters 
(i.e., superclusters) which have more than one neighbor clusters within the 
FoF linkage length, $b\bar{l}$ where $\bar{l}$ is the mean cluster separation. 
The {\it isolated clusters} are then identified as those clusters which do not 
belong to any superclusters. Note that the degree of their isolation depends on 
the value of $b$: The larger the value of $b$ is, the more isolated they are.
Here we consider the extreme case in which the isolated clusters experience no 
disturbance from the surrounding large-scale structures and the difference 
between their FoF and SO masses is negligible. In this extreme case, the 
value of the diffusion coefficient, $D_{B}$, must vanish since the non-zero value 
of $D_{B}$ indicates the presence of the disturbance from the surroundings and 
the existence of the difference between the FoF and SO masses. 

\begin{figure}
\begin{center}
\plotone{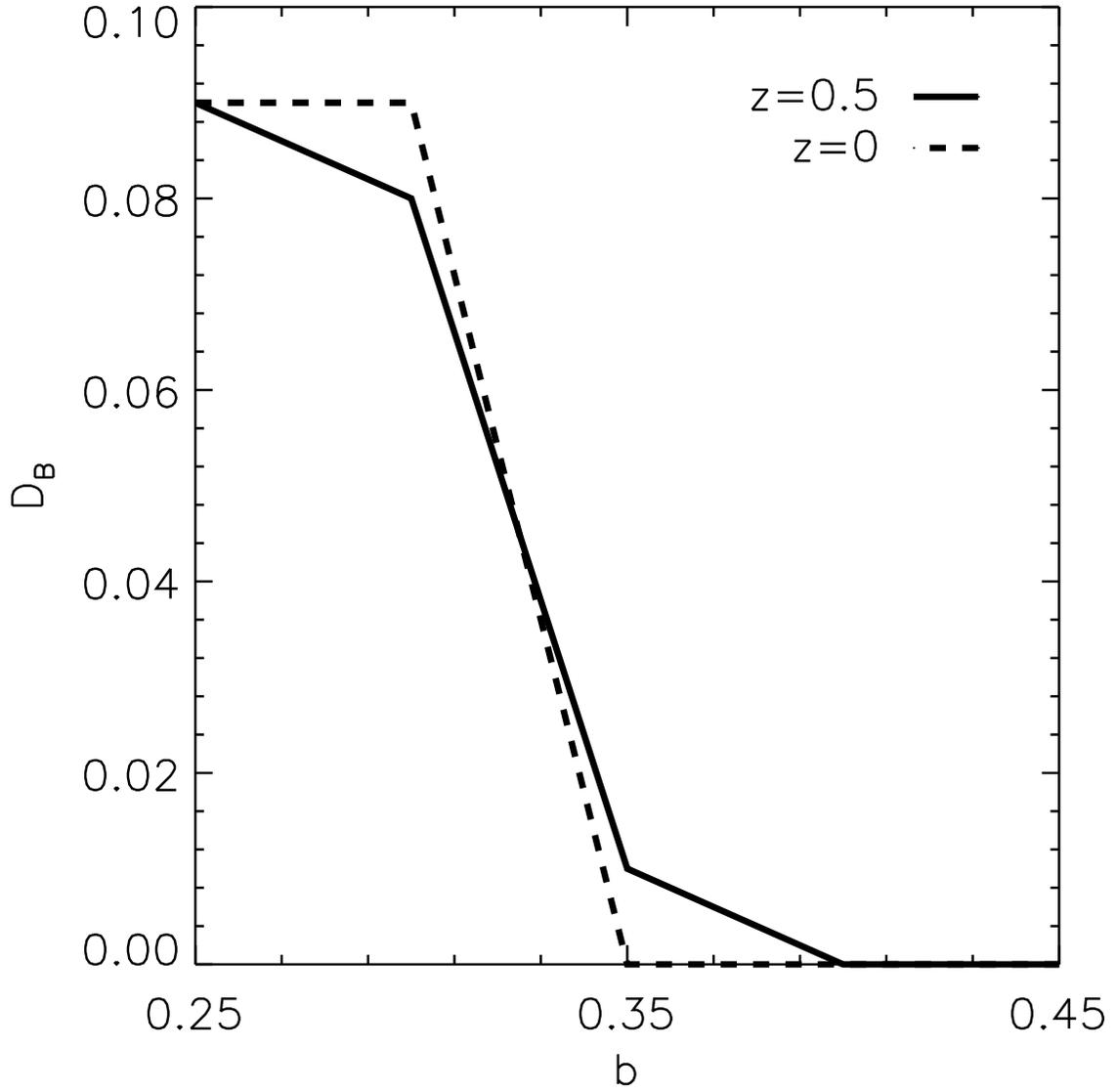}
\caption{Best-fit value of the diffusion coefficient, $D_{B}$, versus the 
linkage parameter, $b$,  used in the FoF algorithm to classify the clusters into 
the isolated and the non-isolated one at $z=0$ and $0.5$ as dashed and solid 
lines, respectively. The result at $z=1$ is very similar to the result at $z=0.5$. }
\label{fig:bDB}
\end{center}
\end{figure}

The ad-hoc value of the linkage length parameter for this extreme case is 
determined to be $b=0.4$ by the following procedures. We first investigate how 
the best-fit value of $D_{B}$ changes as the value of $b$ varies from 
$0.25$ to $0.45$ in the FoF algorithm applied to the MICE cluster catalogs. 
For each case of $b$, we select the isolated clusters 
from the MICE cluster catalog and obtain the numerical mass function of the 
isolated clusters. Then, we fit the numerical result to the analytic CA11 formula 
(Eq.[\ref{eqn:mft}]-[\ref{eqn:f1b2}]) to determine the best-fit 
values of $D_{B}$ and $\beta$ with the help of the $\chi^{2}$ statistics.
Figure \ref{fig:bDB} plots the best-fit value of $D_{B}$ versus the 
linkage length parameter $b$. As can be seen, the best-fit value of $D_{B}$ 
drops to zero when the linkage length parameter $b$ reaches up to $0.4$. 
Table \ref{tab:best2} lists the best-fit values of $D_{B}$ and $\beta$ for 
the mass function of the isolated clusters when the ad-hoc value of the 
linkage length parameter is set at $b=0.4$. 

Now, we write the total mass function of all cluster halos, $dN_{\rm T}/d\ln M$, 
as the sum of the mass function of the isolated cluster halos, 
$dN_{\rm I}/d\ln M$ and that of the non-isolated cluster halos, 
$dN_{\rm NI}/d\ln M$:
\begin{equation}
\label{eqn:sep_t}
\frac{dN_{T}(M, z)}{d\ln M}=\frac{dN_{\rm I}(M, z)}{d\ln M} + 
\frac{dN_{\rm NI}(M, z)}{d\ln M}.
\end{equation}
Putting $D_{B}=0$ in the CA11 formula, we express the mass function of the 
isolated clusters as
\begin{equation}
\label{eqn:separ}
\frac{dN_{I}(M, z)}{d\ln M} = \frac{\bar{\rho}}{M}\frac{d\ln\sigma^{-1}}
{d\ln M}f[\sigma;\beta, D_{B}=0].
\end{equation}
Here the multiplicity function of the isolated clusters which is characterized by 
one coefficient $\beta$ is written at second order as 
\begin{equation}
\label{eqn:ca10_mul_I}
f(\sigma ; D_{B}=0,\beta) \approx f^{(0)}(\sigma) + 
f^{(1)}_{\beta=0}(\sigma) + 
f^{(1)}_{\beta}(\sigma) + 
f^{(1)}_{\beta^2}(\sigma),
\end{equation}
where
\begin{eqnarray}
\label{f0_I}
f^{(0)}(\sigma)&=&\frac{\delta_{sc}}{\sigma}
\sqrt{\frac{2}{\pi}}\,e^{-\frac{(\delta_{sc}+\beta\sigma^2)^2}{2\sigma^2}},\\
\label{f1b0_I}
f^{(1)}_{\beta=0}(\sigma)&=&-\kappa\frac{\delta_{sc}}{\sigma}
\sqrt{\frac{2}{\pi}}\left[e^{-\frac{\delta_{sc}^2}{2\sigma^2}}
-\frac{1}{2}\Gamma\left(0,x\right)\right],\\
\label{f1b1_I}
f^{(1)}_{\beta}(\sigma)&=&
-\beta\,\delta_{sc}\left[f^{(1)}_{\beta=0}(\sigma)+\kappa\,\textrm{Erfc}
\left(x\right)\right],\\
\label{f1b2_I}
f^{(1)}_{\beta^2}(\sigma)&=&\beta^{2}\delta^{2}_{sc}\kappa
\biggl\{\textrm{Erfc}\left(x\right)+
\frac{\sigma}{\delta_{sc}}\sqrt{\frac{1}{2\pi}}\biggl[e^{-\frac{\delta_{sc}^2}
{2\sigma^2}}\left(\frac{1}{2}-2x\right)+\frac{3}{4}\frac{\delta_{sc}^2}
{\sigma^2}\Gamma\left(0,x\right)\biggr]\biggr\}.
\end{eqnarray}

\begin{figure}
\begin{center}
\plotone{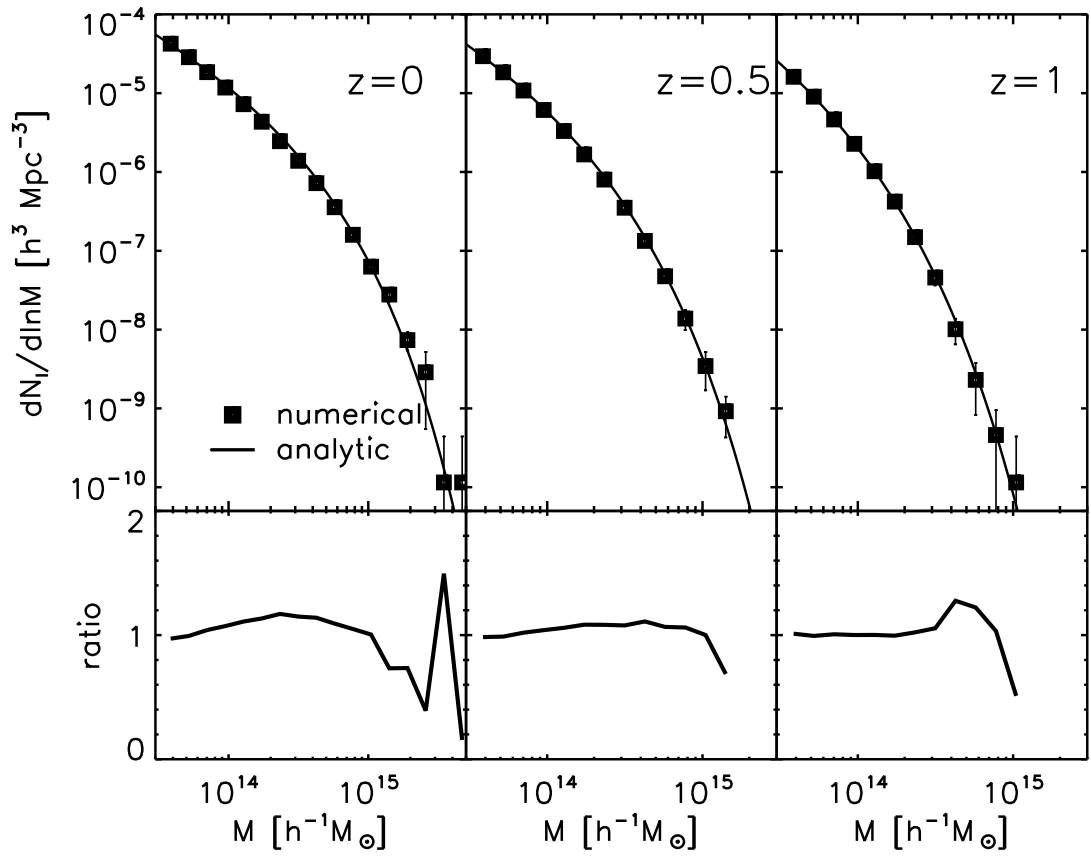}
\caption{Same as Figure \ref{fig:dndm} but for the isolated cluster halos.}
\label{fig:dnIdm}
\end{center}
\end{figure}

Figure \ref{fig:dnIdm} plots the analytical mass functions of the isolated 
clusters (solid lines) with the best-fit values of $\beta$ and compared them with 
the numerical results from the MICE simulations (square dots) at $z=0,\ 0.5$ and 
$1$ in the top-left, top-middle and top-right panel, respectively. 
As can be seen, the analytical mass functions of the isolated clusters agree 
quite well with the numerical results at all three redshifts. 
The ratios of the analytic CA11 formula to the numerical results 
are plotted in the bottom panels. Although the agreements between the analytical 
and the numerical results for the case of the isolated clusters are not so 
excellent as for the case of all cluster halos (see Figure \ref{fig:dndm}), 
the ratio at each redshift is still quite close to unity: the discrepancy is 
less than $20\%$ except in the high-mass limit where the Jackknife errors are 
very large. 

This result indicates that the CA11 formula with $D_{B}=0$ indeed works 
fairly well for the determination of the mass function of the isolated clusters.
It is worth emphasizing here that this analytical mass function of the isolated 
clusters (Eq.~[\ref{eqn:separ}]) has only one coefficient $\beta$ and there is 
no fitting normalization constant. The amplitude of $dN_{\rm I}/d\ln M$ is 
found to automatically match the numerical result when the best-fit value of 
$\beta$ is put into the CA formula. 

\begin{figure}
\begin{center}
\plotone{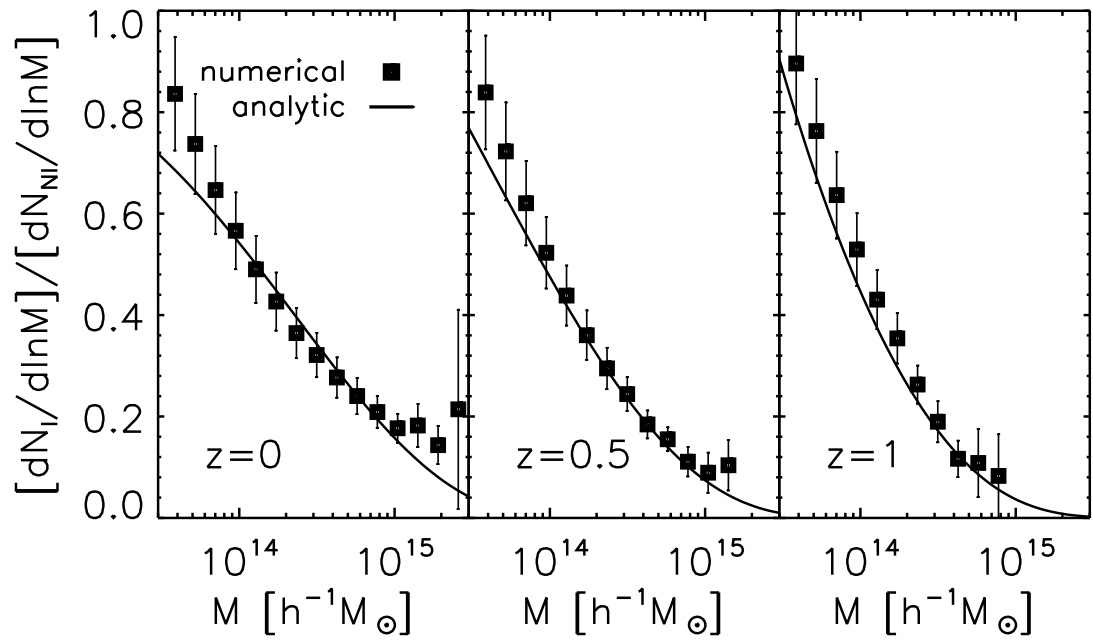}
\caption{Ratio of the mass function of the isolated clusters 
to that of the non-isolated clusters at $z=0,\ 0.5$ and $1$ in the 
left, middle and right panel, respectively.}
\label{fig:ratio}
\end{center}
\end{figure}

Figure \ref{fig:ratio} plots the ratio of the mass function of the isolated 
clusters to that of the non-isolated clusters at $z=0,\ 0.5$ and $1$ in the 
left, middle and right panel, respectively. The errors are calculated as 
one $\sigma$ scatter of the ratio among the eight Jackknife resamples. 
As can be seen, the numerical result (square dots) agrees fairly well with the 
analytic prediction (solid line) based on the CA11 formula at each redshift. 
We believe that it would reduce the existing discrepancy between the analytical and 
numerical results even more to include the higher order terms in the calculation of the 
multiplicity function, $f(\sigma ; D_{B},\beta)$. Throughout this paper, however, 
we consider only the second order approximation of $f(\sigma ; D_{B},\beta)$. 
Note that the ratio drops with mass more rapidly at higher redshifts, indicating 
that at high redshifts most of the very massive clusters have stronger 
tendency to form in the highly overdense regions than at present epoch.

\section{DEPENDENCE ON THE DARK ENERGY EQUATION OF STATE}

In a $\Lambda$CDM cosmology the dark energy equation of state (defined as the 
ratio of the pressure density to the energy density) is a perfect constant given 
as $w\equiv P_{\Lambda}/\rho_{\Lambda}=-1$. Whereas in dynamic dark energy 

models the value of $w$ may vary with time and can deviate from $-1$ at $z=0$ 
\citep{WS98}. This difference in the dark energy equation of state results in the 
different functional shape of $D(z)$ \citep{cp01,lin03,bas03,per05}. 

The evolution of the abundance of galaxy clusters has been regarded as 
one of the most powerful probes of $w(z)$ 
\citep[e.g.,][]{WS98,hai-etal01,wel-etal02}. 
The most serious systematics in constraining $w(z)$ with the evolution 
of the abundance of the galaxy clusters may come from the mass assignment of the 
high-$z$ clusters. Several statistical methods have been suggested so far to 
overcome the systematics but none of them have yet to be fully satisfactory 
\citep[e.g.,][and references therein]{rykoff-etal08,CE10,stanek-etal10}. 
See also Allen et al.(2011) for the latest review.
We suggest here that the relative abundance of the galaxy clusters at a given 
epoch provides a complimentary probe of the dark energy equation of state. 

The relative  abundance of the isolated clusters is defined as the 
ratio of the cumulative mass function of the isolated clusters to that of the 
non-isolated clusters:
\begin{equation}
\label{eqn:rel_abun}
\xi_{\rm I}(M_{c};z)\equiv \left[\int_{M_{c}}^{\infty}d\ln M
\frac{dN_{\rm I}(M, z)}{d\ln M}\right]\Bigg{/}\left[\int_{M_{c}}^{\infty}d\ln M
\frac{dN_{\rm NI}(M, z)}{d\ln M}\right],
\end{equation}
where the mass function of the non-isolated cluster $dN_{\rm NI}/d\ln M$ is 
obtained by $dN_{\rm T}/d\ln M - dN_{\rm I}/d\ln M$.
To demonstrate how the relative abundance of the isolated clusters 
$\xi_{\rm I}$ changes with the dark energy equation state $w$, we consider 
the specific case where the dark energy equation of state is redshift dependent 
as $w(z)=w_{0}+w_{1}z/(1+z)$ \citep{cp01,lin03}. 
\citet{bas03} and \citet{per05} showed that for these dynamic dark energy models 
the linear growth factor can be approximated as:  
\begin{equation}
D(z)=\frac{5\Omega_{mz}}{2(z+1)}
\left[\Omega^{\alpha}_{mz}-\Omega_{Qz} + 
\left(1+\frac{\Omega_{mz}}{2}\right)
\left(1+{\cal A}\Omega_{Qz}\right)\right]^{-1},
\label{eqn:D_qcdm}
\end{equation}
where
\begin{eqnarray}
\label{eqn:alp}
\alpha &=&\frac{3}{5-2/(1-w)}+\frac{3}{125}
\frac{(1-w)(1-3w/2)}{(1-6w/5)^{3}}[1-\Omega_{mz}],\\
\label{eqn:cal}
{\cal A}&=&-\frac{0.28}{w+0.08}-0.3.
\end{eqnarray}
Here $\Omega_{mz}$ and $\Omega_{Qz}$ represent the matter density and 
dark energy density parameters at $z$, respectively, related to their present values, 
$\Omega_{m}$ and $\Omega_{Q}$ as   
\begin{equation}
\Omega_{mz}=\frac{\Omega_{m}(1+z)^{3}}{E^{2}(z)}, \quad
\Omega_{Qz}=\frac{\Omega_{Q}}{E^{2}(z)(1+z)^{f(z)}}.
\label{eqn:ome} 
\end{equation}
where
\begin{eqnarray}
\label{eqn:e2}
E^{2}(z)&=&\Omega_{m}(1+z)^{3}+\Omega_{Q}(1+z)^{-f(z)},\\
\label{eqn:fz}
f(z) &=& -3(1+w_{0})-\frac{3w_{1}}{2\ln(1+z)}.
\end{eqnarray}

\begin{figure}
\begin{center}
\plotone{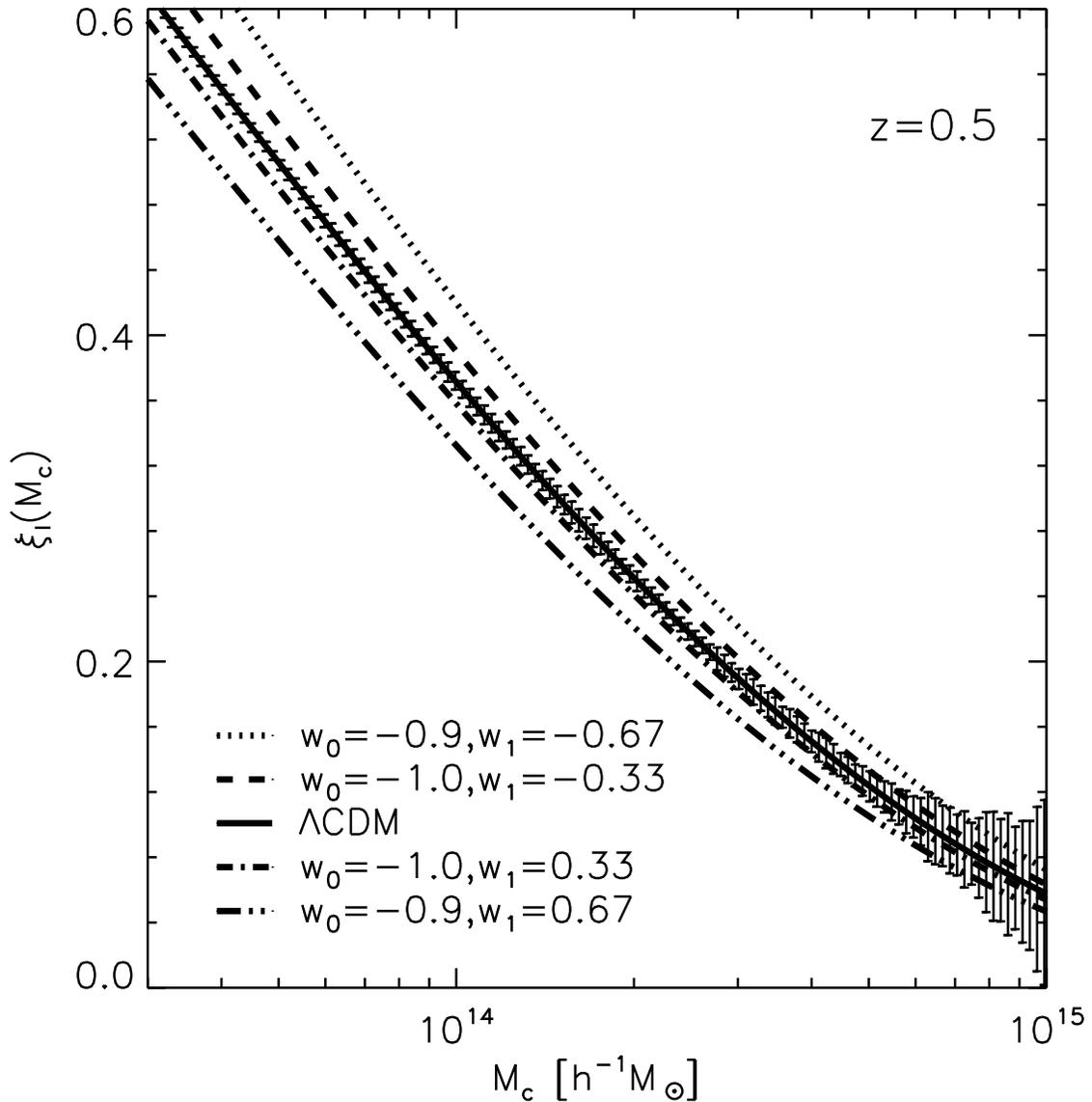}
\caption{Relative abundance of the isolated clusters 
(Eq.~[\ref{eqn:rel_abun}]) for five different cases of the dark 
energy equation of state: $w(z)=w_{0}+w_{1}z/(1+z)$. The errors for the 
$\Lambda$CDM-case are obtained as the one $\sigma$ scatter among the 
eight Jackknife resamples.}
\label{fig:niw}
\end{center}
\end{figure}

We first calculate $\xi_{\rm I}$ for the $\Lambda$CDM case and then repeat the 
calculation of $\xi_{\rm I}$ for the four different dynamic dark energy models 
at $z=0.5$.   
Figure \ref{fig:niw} plots the relative abundances of the isolated clusters for 
the dynamic dark energy models and compare them with the $\Lambda$CDM case.
The errors for the $\Lambda$CDM case are calculated as one $\sigma$ scatter of 
$\xi_{\rm I}$ among the eight Jackknife resamples from the MICE cluster catalog. 
As can be seen, the five different dark energy models predict different relative 
abundances of the isolated clusters. 
For the case of $w_{1}>0$ ($w_{1}<0)$, the relative abundance of the isolated 
clusters has lower (higher) amplitude than for the $\Lambda$CDM case. 
This can be explained by the following logic. In a dark energy model with 
$w_{1}>0$ ($w_{1}<0)$ the largest-scale powers are higher (lower) than in the 
$\Lambda$CDM case and thus boost the merging of the isolated clusters into 
the superclusters, reducing the relative abundance of the isolated clusters.

Furthermore, the differences in $\xi_{\rm I}$ between each dynamic dark energy 
case and the $\Lambda$CDM case are larger than the statistical errors 
calculated through the Jackknife resampling. If a future cluster survey 
can find as many clusters as the MICE simulations, the sensitivity shown in 
Figure \ref{fig:niw} suggests that the relative abundance of the isolated 
clusters will  be a powerful probe of the dark energy equation of state in practice. 

\begin{figure}
\begin{center}
\plotone{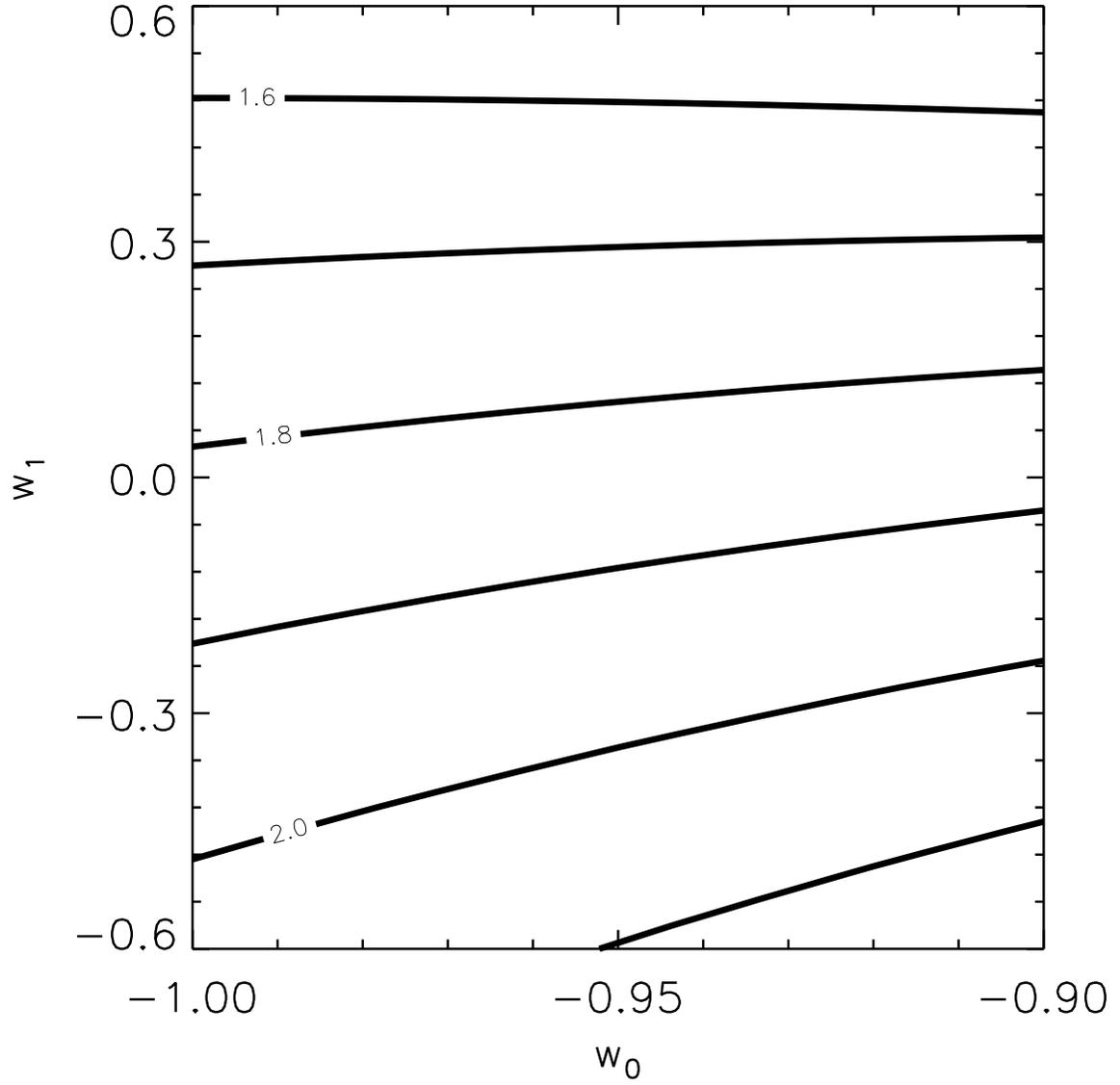}
\caption{Contour plot for the relative abundance of the isolated clusters at 
$z=0.5$ in the $w_{0}$-$w_{1}$ plane.}
\label{fig:con_ww}
\end{center}
\end{figure}

We examine the degeneracy between $w_{0}$ and $w_{1}$.  Varying the value 
of $w_{0}$ from $-1$ to $-0.9$ and the value of $w_{1}$ from $-0.6$ to $0.6$, 
we repeatedly calculate $\xi_{\rm I}$ at $z=0.5$, setting the 
cutoff mass scale at $M_{c}=3.35\times 10^{13}\,h^{-1}M_{\odot}$ 
(the lowest cluster mass in the MICE catalog).
The contour curves of $\xi_{\rm I}$ in the $w_{0}$-$w_{1}$ plane 
are plotted in Figure \ref{fig:con_ww}. When $w_{0}$ is fixed, the relative 
abundance of the isolated clusters decreases as the value of $w_{1}$ 
increases, which is consistent with Figure \ref{fig:niw}. 

\begin{figure}
\begin{center}
\plotone{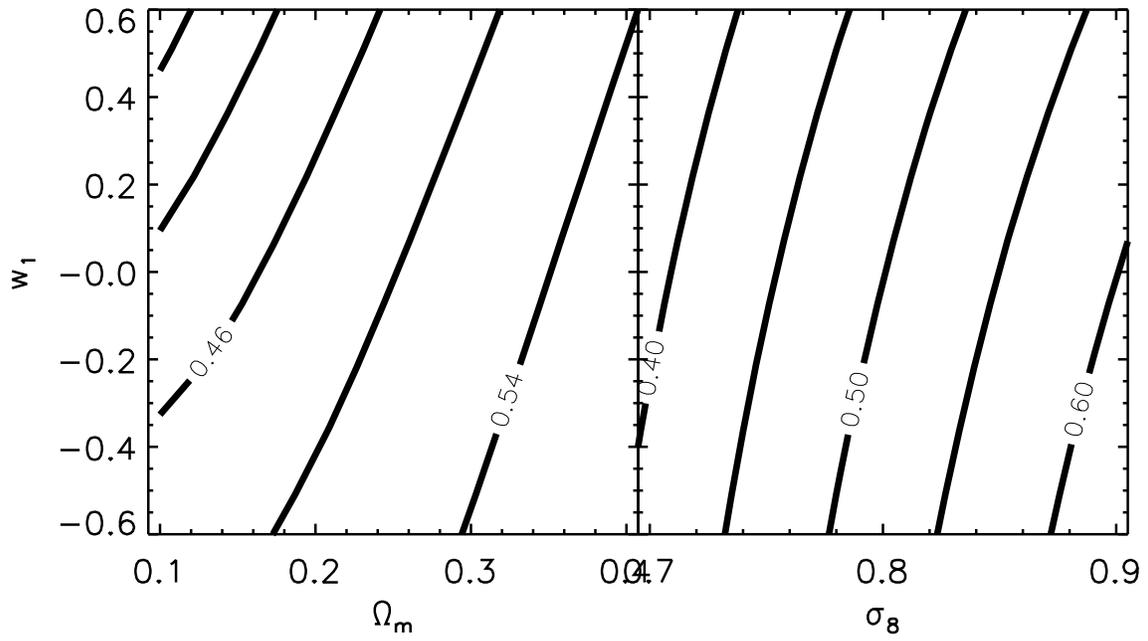}
\caption{Contour plot for the relative abundance of the isolated clusters at 
$z=0.5$ in the $\Omega_{m}$-$w_{1}$ plane (left panel) and in the 
$\sigma_{8}$-$w_{1}$ plane (right panel).}
\label{fig:con_para}
\end{center}
\end{figure}

We also examine the degeneracy between $\Omega_{m}$ and $w_{1}$ and between 
$\sigma_{8}$ and $w_{1}$ at $z=0.5$, which are plotted in the left 
and right panels of Figure \ref{fig:con_para}, respectively. For these plots the 
value of $w_{0}$ is set at $-1$. As can be seen, when $w_{1}$ is fixed, the 
relative abundance of the isolated clusters,  $\xi_{\rm I}$, increases as the 
value of $\Omega_{m}$ increases.  This implies that if there is stronger 
gravitational effect due to the larger amount of dark matter, the formation of a 
cluster can occur more easily even in isolated low-density environments. 
For a given value of $\xi_{\rm I}$, $w_{1}$ increases with $\Omega_{m}$.
A similar trend is found in the $\sigma_{8}$-$w_{1}$ degeneracy curves. 
When $w_{1}$ is fixed, $\xi_{\rm I}$ increases as the value of $\sigma_{8}$ 
increases. The overall high density powers make it less hard for 
a cluster to form in the isolated low-density environments. 

It is worth mentioning here that the degeneracy trends shown in Figures 
\ref{fig:con_ww}-\ref{fig:con_para} are different from those obtained from the 
total cluster abundance \citep[see][]{WS98,wel-etal02}. The increase of 
$w_{1}$ has the same effect on the cluster abundance as the increase of 
$\Omega_{m}$ and $\sigma_{8}$: 
the cluster abundance increases as the three parameters increase.
In contrast, when the relative abundance of isolated clusters is used, the 
increase of $w_{1}$ has the opposite effect: The relative abundance of the 
isolated clusters increases as $w_{1}$ decreases and as $\Omega_{m}$ and 
$\sigma_{8}$ increase. This result implies that the relative abundance of 
the isolated clusters may be helpful to break the degeneracy between $w_{1}$ 
and the other key cosmological parameters.
 
\section{SUMMARY AND DISCUSSION}

We have derived the mass function of the isolated clusters in the frame of 
the recently developed Corasaniti-Achitouv formalism, assuming that for the 
case of the isolated clusters there is no disturbance from the surroundings 
and no ambiguity in the mass determination.  The numerical results from the 
MICE simulations have been used to determine empirically the value of the 
drifting average coefficient, $\beta$, which quantifies the degree of the 
deviation of the collapse barrier for the formation of the isolated clusters 
from the standard average value. 

Extrapolating the validity of the Corasaniti-Achitouv formalism to the 
dynamic dark energy models and using our analytic result on the mass function 
of the isolated clusters,  we have shown that the relative abundance of the 
isolated clusters,  $\xi_{\rm I}$, defined as the ratio of the cumulative mass 
function of the isolated clusters to  that of the non-isolated clusters at a 
given epoch, depends sensitively on the dark energy equation of state. 
This result proves our theoretical concept that the relative abundance of the 
isolated clusters is in principle a powerful probe of dark energy.

Yet, several additional works have to be done before constraining the dark 
energy equation of state with $\xi_{\rm I}$.  
First, it will be necessary to derive the functional form of the drifting 
average coefficient, $\beta(z)$, for the isolated and non-isolated 
cases, separately.
Second,  it has to be confirmed if the Corasaniti-Achitouv formalism 
indeed works not only in a $\Lambda$CDM cosmology but also in dynamic 
dark energy models and how the drifting average coefficient changes 
with the background cosmology.
Third, it has to be examined whether or not the higher-order perturbation 
terms in the Corasaniti-Achitouv formalism have any non-negligible effect 
on the relative abundance of the isolated clusters. We plan to conduct 
these works and wish to report the results elsewhere in the future.

We would also like to mention that our analytic model for the mass function 
of the isolated clusters will be useful not only as a probe of dark energy 
equation of state but also for predicting more accurately the cluster-related 
statistics since the isolated massive clusters may differ from the 
non-isolated ones in their physical properties such as average shape, dynamical 
state, gas entropy profile and etc (in private communication with E.Komatsu).

\acknowledgments

I thank a referee for helpful suggestions and E. Komatsu for stimulating discussion.
I acknowledge the use of data from the MICE simulations that are 
publicly available at http://www.ice.cat/mice.  I also acknowledge the 
financial support from the National Research Foundation of Korea (NRF) 
grant funded by the Korea government (MEST, No.2011-0007819) and from the 
National Research Foundation of Korea to the Center for Galaxy Evolution 
Research.

\clearpage
\clearpage

\clearpage
\clearpage
\clearpage
\clearpage
\clearpage
\clearpage
\begin{deluxetable}{cccc}
\tablewidth{0pt}
\setlength{\tabcolsep}{5mm}
\tablecaption{Redshift, total number of clusters, and best-fit values 
of the two coefficients in the CA11 formalism.}
\tablehead{z & $N_{\rm T}$ & $D_{B}$ & $\beta$}
\startdata
$0$  & $2819031$ & $0.38$ & $0.107$ \\
$0.5$ & $1684018$ & $0.38$ & $0.095$\\
$1$ & $749614$ & $0.38$ & $0.078$
\enddata
\label{tab:best1}
\end{deluxetable}
\clearpage
\begin{deluxetable}{cccc}
\tablewidth{0pt}
\setlength{\tabcolsep}{5mm}
\tablecaption{Redshift, number of isolated clusters, and best-fit values 
of the two coefficients in the CA formalism}
\tablehead{z & $N_{\rm I}$ & $D_{B}$ & $\beta$}
\startdata
$0$  & $1334200$ & $0.0$ & $0.34$ \\
$0.5$ & $782254$ & $0.0$ & $0.20$\\
$1$ & $358073$ & $0.0$ & $0.08$
\enddata
\label{tab:best2}
\end{deluxetable}

\clearpage
\begin{thebibliography}{}
\bibitem[Allen et al.(2011)]{allen-etal11} Allen, S.~W., Evrard, 
A.~E., \& Mantz, A.~B.\ 2011, arXiv:1103.4829 
\bibitem[Audit et al.(1997)]{audit-etal97}
Audit, E., Teyssier, R., \& Alimi, J.~M.\ 1997, \aap, 325, 439
\bibitem[Basilakos(2003)]{bas03}
Basilakos, S. 2003, \apj, 590, 636
\bibitem[Basilakos et al.(2010)]{bas-etal10} Basilakos, S., 
Plionis, M., \& Lima, J.~A.~S.\ 2010, \prd, 82, 083517 
\bibitem[Bardeen et al.(1986)]{bbks86}
Bardeen, J.~M., Bond, J.~R., Kaiser, N., \& Szalay, A.~S.\ 1986, \apj, 304, 15
\bibitem[Bernardeau(1994)]{bernardeau94} 
Bernardeau, F.\ 1994, \apj, 427, 51 
\bibitem[Bond et al.(1991)]{bond-etal91}
Bond, J.~R., Cole, S., Efstathious, G., \& Kaiser N.\ 1991, \apj, 379, 440
\bibitem[Bond \& Myers(1996)]{BM96} 
Bond, J.~R., \& Myers, S.~T.\ 1996, \apjs, 103, 1 
\bibitem[Chevallier \& Polarski(2001)]{cp01}
Chevallier, M. \& Polarski, D. 2001, Int. J. Mod. Phys. D, 10, 213
\bibitem[Corasaniti \& Achitouv(2011a)]{CA11a} 
Corasaniti, P.~S., \& Achitouv, I.\ 2011, Physical Review Letters, 106, 241302 
\bibitem[Corasaniti \& Achitouv(2011b)]{CA11b} 
Corasaniti, P.~S., \& Achitouv, I.\ 2011, arXiv:1107.1251 
\bibitem[Crocce et al.(2010)]{mice10} Crocce, M., Fosalba, P., 
Castander, F.~J., \& Gazta{\~n}aga, E.\ 2010, \mnras, 403, 1353 
\bibitem[Cunha \& Evrard(2010)]{CE10} 
Cunha, C.~E., \& Evrard, A.~E.\ 2010, \prd, 81, 083509 
\bibitem[Davis et al.(1985)]{fof85}
Davis, M., Efstathiou, G., Frenk, C.~S., \& White, S.~D.~M.\ 1985,
\apj, 292, 371
\bibitem[Einasto et al.(2007)]{einasto-etal07} 
Einasto, J., et al.\ 2007, \aap, 462, 811 
\bibitem[Eke et al.(1996)]{eke-etal96}
Eke, V.~R., Cole, S., \& Frenk, C.~S.\ 1996, \mnras, 282, 263
\bibitem[Gunn \& Gott(1972)]{GG72} 
Gunn, J.~E., \& Gott, J.~R., III 1972, \apj, 176, 1 
\bibitem[Haiman et al.(2001)]{hai-etal01}
Haiman, Z., Mohr, J.J. \& Holder, G.P. 2001, \apj, 553, 545
\bibitem[Jedamzik(1995)]{jedamzik95} 
Jedamzik, K.\ 1995, \apj, 448, 1 
\bibitem[Jenkins et al.(2001)]{jenkins-etal01}
Jenkins, A., et al.\ 2001, \mnras, 321, 372
\bibitem[Jones et al.(1993)]{jones-etal93} 
Jones, M., et al.\ 1993, \nat, 365, 320  
\bibitem[Kasun \& Evrard(2005)]{KE05} 
Kasun, S.~F., \& Evrard, A.~E.\ 2005, \apj, 629, 781 
\bibitem[Komatsu et al.(2011)]{wmap7} Komatsu, E., et al.\ 
2011, \apjs, 192, 18 
\bibitem[Lahav et al.(1991)]{lah-etal91} 
Lahav, O., Lilje, P.~B., Primack, J.~R., \& Rees, M.~J., 1991, 
MNRAS, 251, 128 
\bibitem[Lee \& Shandarin(1998)]{LS98}
Lee, J. \& Shandarin, S.~F.\ 1998, \apj, 500, 14
\bibitem[Lee \& Evrard(2007)]{LE07} 
Lee, J., \& Evrard, A.~E.\ 2007, \apj, 657, 30 
\bibitem[Linder(2003)]{lin03}
Linder, E. 2003, \prl, 90, 091301
\bibitem[Linder(2005)]{linder05} 
Linder, E.~V.\ 2005, \prd, 72, 043529 
\bibitem[Lue et al.(2004)]{lue-etal04} 
Lue, A., Scoccimarro, R., \& Starkman, G.~D.\ 2004, \prd, 69, 124015 
\bibitem[Maggiore \& Riotto(2010a)]{MR10a} 
Maggiore, M., \& Riotto, A.\ 2010, \apj, 711, 907
\bibitem[Maggiore \& Riotto(2010b)]{MR10b} 
Maggiore, M., \& Riotto, A.\ 2010, \apj, 717, 515 
\bibitem[Maggiore \& Riotto(2010c)]{MR10c} 
Maggiore, M., \& Riotto, A.\ 2010, \apj, 717, 526
\bibitem[Monaco(1997)]{monaco97}
Monaco, P.\ 1997, \mnras, 290, 439  
\bibitem[Percival(2005)]{per05}
Percival, W. J. 2005, \aap, 819, 830
\bibitem[Press \& Schechter(1974)]{PS74} 
Press, W.~H., \& Schechter, P.\ 1974, \apj, 187, 425 
\bibitem[Reed et al.(2003)]{reed-etal03} Reed, D., Gardner, J., 
Quinn, T., Stadel, J., Fardal, M., Lake, G., 
\& Governato, F.\ 2003, \mnras, 346, 565 
\bibitem[Robertson et al.(2009)]{robertson-etal09} 
Robertson, B.~E., Kravtsov, A.~V., Tinker, J., \& 
Zentner, A.~R.\ 2009, \apj, 696, 636 
\bibitem[Rykoff et al.(2008)]{rykoff-etal08} 
Rykoff, E.~S., et al.\ 2008, \mnras, 387, L28 
\bibitem[Sheth \& Tormen(1999)]{ST99} 
Sheth, R.~K., \& Tormen, G.\ 1999, \mnras, 308, 119 
\bibitem[Sheth et al.(2001)]{SMT01} 
Sheth, R.~K., Mo, H., \& Tormen, G.\ 2001, \mnras, 323, 1 
\bibitem[Springel et al.(2001)]{springel-etal01} Springel, V., White, 
M., \& Hernquist, L.\ 2001, \apj, 549, 681  
\bibitem[Springel et al.(2005)]{millennium05}
Springel, V., et al.\ 2005, \nat, 435, 629
\bibitem[Stanek et al.(2010)]{stanek-etal10} Stanek, R., Rasia, E., 
Evrard, A.~E., Pearce, F., \& Gazzola, L.\ 2010, \apj, 715, 1508 
\bibitem[Tinker et al.(2008)]{tinker-etal08}
Tinker, J.~L., et al.\ 2008, \apj, 688, 709
\bibitem[Wang \& Steinhardt(1998)]{WS98}
Wang, L. \& Steinhardt, P. J. 1998, \apj, 508, 483
\bibitem[Warren et al.(2006)]{warren-etal06}
Warren, M.~S., Abazajian, K., Holz, D.~E., \& Teodoro, L.\ 2006, \apj, 646, 881
\bibitem[Weller et al.(2002)]{wel-etal02}
Weller, J., Battye, R. A., \& Kneissl, R. 2002, \prl, 88, 231301
\bibitem[Wray et al.(2006)]{wray-etal06} Wray, J.~J., Bahcall, 
N.~A., Bode, P., Boettiger, C., \& Hopkins, P.~F.\ 2006, \apj, 652, 907 
\end{thebibliography}
\end{document}